\documentclass[10pt]{amsart}
\usepackage{amsmath,amsfonts,amssymb,amsthm,amscd}
\newtheorem{thm}{Theorem}
\newtheorem{lem}[thm]{Lemma}
\newtheorem{defn}[thm]{Definition}
\newtheorem{prop}[thm]{Proposition}

\def\res{\mathop{\rm res}\limits}
\begin{document}

\title[Jacobi groups and Elliptic solutions of the WDVV equations]{Duality for Jacobi group orbit spaces and elliptic solutions of the WDVV equations}

\author{Andrew Riley$^{(1)}$ and Ian A. B. Strachan$^{(2)}$}

\date{9$^{\rm th}$ November 2005}
\address[1]{Department of Mathematics\\ University of Hull\\
Hull HU6 7RX\\ U.K.}

\email{a.riley@math.hull.ac.uk}

\address[2]{Department of Mathematics\\ University of Glasgow\\ Glasgow G12 8QQ\\ U.K.}

\email{i.strachan@maths.gla.ac.uk}

%\subjclass{11F55, 53B50, 53D45}

\begin{abstract}
From any given Frobenius manifold one may construct a so-called \lq dual\rq~structure which, while not satisfying the full axioms of a Frobenius manifold, shares many of its essential features, such as the existence of a prepotential
satisfying the WDVV equations of associativity. Jacobi group orbit spaces naturally carry the structures of a
Frobenius manifold and hence there exists a dual prepotential. In this paper this dual prepotential is constructed
and expressed in terms of the elliptic polylogarithm function of Beilinson and Levin.
\end{abstract}

\maketitle

\section{Introduction}

The Witten-Dijkgraaf-Verlinde-Verlinde (or WDVV) equation
\[
\frac{\partial^3F}{\partial t^\alpha \partial t^\beta \partial t^\lambda}\eta^{\lambda\mu}
\frac{\partial^3F}{\partial t^\mu \partial t^\gamma \partial t^\delta}-
\frac{\partial^3F}{\partial t^\delta \partial t^\beta \partial t^\lambda}\eta^{\lambda\mu}
\frac{\partial^3F}{\partial t^\mu \partial t^\gamma \partial t^\alpha}=0\,,\quad \alpha\,,\beta\,,\gamma\,,\delta=1\,\ldots\,,n
\]
originally appeared in the theory of 2D topological quantum field theories, with the geometric
structure behind this equation being formalized by Dubrovin into the concept of a Frobenius manifold.

\begin{defn}
An algebra $(\mathcal{A},\circ,\eta)$ over $\mathbb{C}$ is a Frobenius algebra if:
\begin{itemize}
\item{} the algebra $\{\mathcal{A},\circ\}$ is commutative, associative with unity $e\,;$
\item{} the multiplication is compatible with a $\mathbb{C}$-valued bilinear, symmetric, nondegenerate
inner product
\[
\eta\,: \,\mathcal{A}\times \mathcal{A}\rightarrow\mathbb{C}
\]
in the sense that
\[
\eta(a \circ b,c) = \eta(a,b\circ c)
\]
for all $a,b,c\in\mathcal{A}\,.$
\end{itemize}
\end{defn}

\noindent With this structure one may defined a Frobenius manifold
\cite{dubrovin1}:
\begin{defn} $(M,\circ,e,\eta,E)$ is a Frobenius manifold if each tangent
space $T_pM$ is a Frobenius algebra varying smoothly over $M$ with the
additional properties:
\begin{itemize}
\item{} the inner product is a flat metric on $M$ (the term \lq metric\rq~will denote a
complex-valued quadratic form on $M$).
\item{} $\nabla e=0$, where $\nabla$ is the Levi-Civita connection of the metric;
\item{} the tensor $(\nabla_W \circ)(X,Y,Z)$ is totally symmetric for all vectors
$W,X,Y,Z \in TM\,;$
\item{} a vector field $E$ exists such that
\[
\nabla(\nabla E) = 0\]
and that the corresponding one-parameter group of diffeomorphisms acts by
conformal transformations of the metric and by rescalings on the Frobenius
algebras $T_pM\,.$
\end{itemize}

\end{defn}
\noindent Since the metric $\eta$ is flat there exists a distinguished coordinate system of flat coordinates
$\{t^\alpha\,,\alpha=1\,,\ldots\,,n\}$ in which the components of the metric are constant. From the various symmetry
properties of tensors $\circ$ and $\nabla\circ$ it then follows that there exists a function $F$, the
prepotential, such that
\begin{eqnarray*}
c_{\alpha\beta\gamma} & = &
\eta\left(\frac{\partial~}{\partial t^\alpha} \circ \frac{\partial~}{\partial t^\beta},\frac{\partial~}{\partial t^\gamma}\right)\,,\\ & = &
\frac{\partial^3F}{\partial t^\alpha \partial t^\beta \partial t^\gamma}\,,
\end{eqnarray*}
and the associativity condition then implies that $F$ satisfies the WDVV-equation.

On such a Frobenius manifold there exists another metric $g$ - the intersection form - defined
by the formula
\[
g^{-1}(x,y) = E(x \circ y)
\]
where $x\,,y\in T^*M$ and the metric $\eta$ has been used to extend the multiplication
from the tangent bundle to the cotangent bundle. This metric is also flat and hence there
exists another distinguished coordinate system $\{p^i\,,i=1\,,\ldots\,,n\}$ in which the
components of the intersection form are constant. Explicitly, the notation
\[
G^{ij} = \frac{\partial p^i}{\partial t^\alpha} \frac{\partial p^j}{\partial t^\beta}g^{\alpha\beta}(t)
\]
will be used to denote the components of the intersection form in its own flat coordinate
system.

Another class of solutions to the WDVV equations (though not satisfying the full axioms
of a Frobenius manifold) appeared in Seiberg-Witten theory \cite{Marsh}. The connection between
these two class has been elucidated by Dubrovin \cite{dubrovin2}. Given a Frobenius manifold $M$
one may define a dual multiplication $\star\,: T{\overset{\star}{M}}\rightarrow T{\overset{\star}{M}}$ by
\[
X \star Y = E^{-1} \circ X \circ Y\,,\qquad\forall\,
X\,,Y\in  T{\overset{\star}{M}}\,,
\]
where $E^{-1}$ is the vector field defined by the
equation $E^{-1}\circ E=e$ and $\overset{\star}{M}=M\backslash\Sigma\,,$ where $\Sigma$ is the
(discriminant) submanifold where $E^{-1}$ is undefined.
This new multiplication is clearly commutative and associative with the Euler field playing the
role of the unity field for the new multiplication. Furthermore the multiplication is
compatible with the intersection form
\[
g(X\star Y,Z)=g(X,Y\star Z)\,.
\]
Further properties are inherited from the original Frobenius structure on $M\,,$ in particular:

\medskip

\bigskip

\begin{thm} There exists a function $\overset{\star}{F}(p)$ such that\footnote{In such formulae, Greek indices are
raised and lowered using the metric $\eta$ and Latin indices using the metric $G\,.$}

\[
\frac{\partial^3 \overset{\star}{F}(p)}{\partial p^i\partial p^j \partial p^k}=G_{ia}G_{jb}
\frac{\partial t^\gamma}{\partial p^k}
\frac{\partial p^a}{\partial t^\alpha}
\frac{\partial p^b}{\partial t^\beta}c^{\alpha\beta}_\gamma(t)
\]
and which satisfies the WDVV-equations in the $\{p^i\}$ coordinates.

\end{thm}

Thus given a specific Frobenius manifold one may construct a \lq dual\rq~solution
to the WDVV-equations by constructing the flat-coordinates of the intersection
form and using the above result.
The aim of this paper is to perform such calculations for
the Frobenius manifolds which appear naturally on the Jacobi group
orbit space
\[
\Omega/J({\mathfrak g})
\]
in the specific case where $W_{\mathfrak g}=A_n\,.$ The dual prepotential
is given in terms
elliptic polylogarithm introduced by Beilinson and Levin \cite{BL,Levin}.
This functions has also appeared in the theory of Frobenius manifolds
and the enumeration of curves by Jacobi forms \cite{Kawai}.
Before this, the similar calculation of the dual prepotential
on the orbit space
\[
\mathbb{C}^n/A_n\,,
\]
will be described, both for completeness and for comparison.

\section{A Landau-Ginzburg/Superpotential construction}

Given a superpotential $\lambda(w)$ the formulae for the various
tensors are given by the following theorem:

\begin{thm}\label{superpotential}

\begin{eqnarray*}
\eta(\partial',\partial{''}) & = & - \sum \res_{d\lambda=0}
\frac{\partial'(\lambda(v) dv) \partial{''}(\lambda(v)
dv)}{d\lambda(v)} \,, \\
c\,(\partial',\partial{''},\partial{'''}) & = & - \sum
\res_{d\lambda=0} \frac{\partial'(\lambda(v) dv)
\partial{''}(\lambda(v)
dv)\partial{'''}(\lambda(v)
dv)}{d\lambda(v)} \,, \\
g(\partial',\partial{''}) & = & - \sum \res_{d\lambda=0}
\frac{\partial'(\log\lambda(v) dv) \partial{''}(\log\lambda(v)
dv)}{d\log\lambda(v)} \,, \\
\overset{\star}{c}\,(\partial',\partial{''},\partial{'''}) & = & - \sum
\res_{d\lambda=0} \frac{\partial'(\log\lambda(v) dv)
\partial{''}(\log\lambda(v)
dv)\partial{'''}(\log\lambda(v) dv)}{d\log\lambda(v)} \,.
\end{eqnarray*}

\end{thm}

\medskip

\noindent The first three formulae appeared in \cite{dubrovin1}
while the last follows immediately from the results in
\cite{dubrovin2}. With these, and the basic result that
\[
\sum \res_{v\in \mathcal{S}} \omega =0
\]
for any meromorphic differential $\omega$ on a compact Riemann
surface $\mathcal{S}\,,$ the calculations reduce to the
calculation of residues. More specifically, the locations of
the zero of $d\lambda=0$ are only known implicitly while the zeros
of $\lambda$ are known explicitly. Thus using this one finds
\[
\sum \res_{d\lambda=0} \omega =-
\sum \res_{\lambda=0} \omega -
\sum \res_{{\rm special~points}} \omega
\]
where the \lq special points\rq~are points like zero or infinity,
where the residues may be easily calculated. Thus the residues, and
hence the various tensors,
in Theorem [\ref{superpotential}] may be calculated very simply.

We first consider the orbit space
$\mathbb{C}^n/A_n$ where the function $\lambda$ is just a
traceless polynomial and the Riemann surface is just the Riemann
sphere before studying the main example of the orbit space
$\Omega/J({\mathfrak g})$ where the function $\lambda$ is elliptic
and the corresponding Riemann surface an elliptic curve.

\section{The orbit space $\mathbb{C}^n/A_n$}

Given a Coxeter group $W$ one may study the algebra of $W$-invariant functions over the
vector space $\mathbb{C}\,.$ The orbit space $\mathbb{C}^n/W$ is in fact a manifold and it was
shown by Dubrovin (and earlier by K.Saito) that this manifold inherits the structure of
a Frobenius manifold, and as such, there exists an associated superpotential construction.
When $W=A_n$ the superpotential is particularly simple:
\[
\lambda(v) = \prod_{i=0}^n (v-z^i)
\]
with the constraint $ \sum_{i=0}^n z^i=0$ which ensure the polynomial $\lambda$ is traceless.
The metric is easily calculated
via the residue theorem:
\[
g=\left.\sum_{i=0}^n (dz^i)^2\right|_{\sum_{j=0}^n z^j=0}\,.
\]
Similarly, the structure constants may be calculated in an identical manner and integrated up to yield
the dual prepotential
\begin{equation}
\overset{\star}{F} = \frac{1}{4} \sum_{i\neq j} (z_i-z_j)^2
\log(z_i-z_j)^2\,. \label{Fan}
\end{equation}
The same result may be obtained via the geometry such Coxeter
orbit spaces. This example may be generalised in a number of
different ways:

\begin{itemize}

\item{} By studying other Coxeter groups. This leads to dual prepotential akin to (\ref{Fan})
but where the sum is over the root systems of the corresponding Coxeter group.

\item{} The orbit space $\mathbb{C}^n/A_n$ is isomorphic to the
genus-zero Hurwitz space $H_{0,n+1}(n+1)\,.$ One may easily
generalize to an arbitrary genus-zero Hurwitz space
$H_{0,N}(k_1\,,\ldots\,,k_l)\,.$ This generalization includes the
special case $H_{0,n+1}(k,n+1-k)$ which corresponds to the orbit
space $\mathbb{C}^n/ {\tilde W}^{(k)}(A_{n-1})$ where ${\tilde
W}^{(k)}(A_{n-1})$ is an extended affine Weyl group corresponding
to the Coxeter group $A_{n-1}\,.$

\item{} More generally still, one may study the induced \lq dual\rq~structure on the
discriminant strata of either Coxeter groups \cite{FV} or Hurwitz spaces.

\end{itemize}

\noindent This last case reduces to using the superpotential
\[
\lambda(v) = \prod_{i=0}^n (v-z^i)^{k_i}\,,\qquad\qquad k_i \in\mathbb{Z}
\]
with constraints
\[
\sum_{i=1}^n k_i z^i=0\,.
\]
The corresponding \lq dual\rq~prepotential is
\[
\overset{\star}{F} = \frac{1}{4} \sum_{i\neq j} k_i k_j (z_i-z_j)^2
\log(z_i-z_j)^2\,
\]
(prepotentials of this form and their connection with
deformed root systems where first studied in \cite{ChalVes}).
Note that while the restriction of the intersection form of a
Frobenius manifold to a discriminant submanifold is flat, the
restriction of the metric $\eta$ to a discriminant submanifold is
not flat. While the tensors induced from $\circ$ and $\nabla
\circ$ are still totally symmetric, the curvature obstructs the
existence of an induced prepotential \cite{iabs}. Hence there is no
prepotential to which the above \lq dual\rq~ prepotential is dual.
This may also be rewritten as a sum over a \lq deformed root system\rq.
This extends the result of Feigin and Veselov \cite{FV} to negative values
of the parameters $k_i\,.$

\section{The orbit space $\Omega/J({A_n})$}

Naively, the Jacobi group orbit spaces may be regarded as an elliptic generalization
of Coxeter group orbit spaces. The full definition will not be required
here and the reader is referred to \cite{B},\cite{EZ} and \cite{wirth} for details. The Jacobi group
$J(\mathfrak g)$ (where $\mathfrak g$ is a complex finite dimensional simple Lie algebra of rank $l$
with Weyl group $W$) acts on the space
\[
\Omega = \mathbb{C}\oplus {\mathfrak h} \oplus \mathbb{H}
\]
where $\mathfrak h$ is the complex Cartan subalgebra of $\mathfrak g$ and $\mathbb{H}$ is the
upper-half-plane, and this leads to the study of invariant functions - the Jacobi forms.
Analogous to the Coxeter case, the orbit space
\[
\Omega/J({\mathfrak g})
\]
is a manifold and carries the structure of a Frobenius manifold. Again, the
case $W=A_n$ (where we abuse notation and write $\Omega/J(A_n)$ for the orbit space)
is particularly simple with the superpotential being given
by
\begin{equation}
\lambda(v) = e^{2\pi i u} \prod_{i=0}^l \frac{ \vartheta_1(v-z_i|\tau)}{\vartheta_1(v|\tau)}
\label{ellipticsuperpotential}
\end{equation}
with the constraint $ \sum_{i=0}^l z^i=0\,$ and $(u\,,{\bf z},\tau) \in\mathbb{C}\oplus {\mathfrak h} \oplus \mathbb{H}\,.$
The details of the construction of this superpotential is given in \cite{B}\,.

\begin{lem}
With the above (\ref{ellipticsuperpotential}) superpotential the intersection form for the orbit space
$\Omega/J(A_l)$ is
\[
g=2\pi^2 du\,d\tau - \left.\sum_{i=0}^l (dz^i)^2\right|_{\sum_{j=0}^n z^j=0}\,
\]
and the dual prepotential is
\[
\overset{\star}F =
{2 \pi i}\,\left[\frac{1}{2}\pi^2\tau u^2 -
u \sum_{i=0}^l (z^i)^2\right]+
\overset{\star}{F}_{quantum}({\bf z}|\tau)\,.
\]
where this function is evaluated on the plane
$\sum z^j=0\,.$
\end{lem}

\begin{proof} The calculation of the intersection form in these coordinate
appears in \cite{B}. The normalization of the $\vartheta_1$-function follows
\cite{W} which accounts for various numerical factors. The main device is the use of the
elliptic connection, which again is described in \cite{B}\,. With the normalizations
used here the elliptic connection is

\[
D^\bullet f(v|\tau) = (\nabla^\bullet_\tau f)(v|\tau) +
\frac{\pi i}{2} \frac{\vartheta_1^{'}(v|\tau)}{\vartheta_1(v|\tau)}f^{'}(v|\tau)
\]
where
\[
(\nabla^\bullet_\tau f)(v|\tau)=(\partial_\tau f)(v|\tau) + \frac{\pi i k}{6}
\frac{\vartheta_1^{'''}(0|\tau)}{\vartheta_1(0|\tau)} f(v|\tau)\,.
\]
This connection takes elliptic modular functions of weight $k$ to elliptic modular functions of weight $(k+2)\,.$

It follows from the simple dependence of $\lambda$ on the coordinate $u$
that
\[
\overset{\star}{c}_{u ij}= 2 \pi \sqrt{-1} \, G_{ij}
\]
and this fixes the above $u$-dependence of the dual prepotential.

\end{proof}

\bigskip

\bigskip

\noindent To calculate the term $\overset{\star}{F}_{quantum}$ it is necessary to calculated the
remaining components of the
tensor $\overset{\star}{c}\,.$ We begin with the components $\overset{\star}{c}_{z_iz_jz_k}\,.$

\begin{prop}
The dual structure functions $\overset{\star}{c}_{z_iz_jz_k}$ are given in terms of the third derivatives
of a function $\overset{\star}{F}_{temp}\,,$
\[
\overset{\star}{c}_{z_iz_jz_k} = \frac{\partial^3 \overset{\star}{F}_{temp}}{\partial z^i\partial z^j \partial z^k}\,,
\qquad\qquad
i,j,k=1\,,\ldots\,,l\,,
\]
where\footnote{Note, $\sideset{}{'}\sum_i$ includes the term $i=0\,.$}
\[
\overset{\star}{F}_{temp} = \frac{1}{8} \sideset{}{'}\sum_{i \neq j} \Lambda_3(z^i-z^j) -
\frac{(l+1)}{4} \sideset{}{'}\sum_i \Lambda_3(z^i)\,.
\]
Hence
\[
\overset{\star}{F}_{quantum} = \overset{\star}{F}_{temp} + \frac{1}{2} \sideset{}{'}\sum _{i,j} A_{ij}(\tau) z^i z^j + \sideset{}{'}\sum_i B_i(\tau) z^i + C(\tau)\,.
\]
\end{prop}

\begin{proof} Using Theorem one may calculate these structure functions in an
exactly analogous manner to the calculation of the $A_n$-structure functions.
For example
\begin{eqnarray*}
\overset{\star}{c}_{z_iz_jz_k} & = & \sum_{r\neq 0} \frac{\vartheta'_1}{\vartheta_1}(z^0-z^r|\tau) -
(l+1) \frac{\vartheta'_1}{\vartheta_1}(z^0|\tau)\\
&&+\frac{\vartheta'_1}{\vartheta_1}(z^0-z^i|\tau)+
\frac{\vartheta'_1}{\vartheta_1}(z^0-z^j|\tau)+
\frac{\vartheta'_1}{\vartheta_1}(z^0-z^k|\tau)
\end{eqnarray*}
and similar formulae may be derived for $\overset{\star}{c}_{z_iz_iz_j}$ and $\overset{\star}{c}_{z_iz_iz_i}\,.$
Using the formulae presented in the appendix one may integrate these equations in terms of
the function $\Lambda_3\,.$ Hence by construction
\[
\frac{\partial^3 ~}{\partial z^i\partial z^j \partial z^k}\left(\overset{\star}{F}_{quantum}-\overset{\star}{F}_{temp}\right)=0\,, \qquad\qquad
i,j,k=1\,,\ldots\,,l\,,
\]
from which the general form of $\overset{\star}{F}_{quantum}$ follows.
\end{proof}

\bigskip

To find the three functions $A_{ij}(\tau)\,,B_i(\tau)$ and $C(\tau)$ requires the calculation
of, respectively, the structure functions
$\overset{\star}{c}_{\tau z_i z_j}\,,\overset{\star}{c}_{\tau\tau z_i}$ and $\overset{\star}{c}_{\tau\tau\tau}\,.$
However, by using the modularity property of these functions at the point ${\bf z}={\bf 0}$ one may
find these three functions without having to calculate these structure functions exactly.

\begin{thm}\label{maintheorem}
\begin{eqnarray*}
\overset{\star}{F}_{quantum}&=&-\frac{1}{8}\sideset{}{'}\sum_{i\neq j}
\left\{\mathcal{L}i_3[q^2,e^{2i(z^i-z^j)}] - \mathcal{L}i_3[q^2,1]\right\}\\
& &+
\frac{(l+1)}{4} \sideset{}{'}\sum_i
\left\{\mathcal{L}i_3[q^2,e^{2iz^i}] - \mathcal{L}i_3[q^2,1]\right\}\,.
\end{eqnarray*}
\end{thm}

\begin{proof} Since
\[
\overset{\star}{c}_{\tau z_i z_j}=
\frac{\partial^3\overset{\star}{F}_{temp}}{\partial\tau\partial z^i\partial z^j}+A_{ij}'(\tau)
\]
and $\overset{\star}{F}_{temp}$ is known explicitly, it follows that
\[
A'_{ij}(\tau) = \overset{\star}{c}_{\tau z_i z_j} ({\bf 0}|\tau)\,.
\]
Similarly
\begin{eqnarray*}
B''_i(\tau) & = &\overset{\star}{c}_{\tau \tau z_i } ({\bf 0}|\tau)\,,\\
C'''(\tau) & = & \overset{\star}{c}_{\tau \tau\tau } ({\bf 0}|\tau)-
\left.\frac{\partial^3\overset{\star}{F}_{temp}}{\partial\tau^3}\right|_{{\bf z}={\bf 0}}\,,\\
& = &\overset{\star}{c}_{\tau \tau\tau }({\bf 0}|\tau)+\frac{i \pi^3}{120} (l+1)(l+2) E_4(\tau)\,.
\end{eqnarray*}
Hence it suffices to calculate the remaining structure functions at the special point ${{\bf z}={\bf 0}}\,.$
Using the quasi-modular property
\[
\frac{\vartheta_1'}{\vartheta_1}\left(\frac{z}{\tau}\right\vert\left. -\frac{1}{\tau}\right)=
\frac{2iz}{\pi} + \tau\frac{\vartheta_1'}{\vartheta_1}(z|\tau)
\]
and its differential consequences one may derive the following modularity properties of the structure functions directly from Theorem [\ref{superpotential}]
{\sl without} having to calculate them explicitly:
\begin{eqnarray*}
\overset{\star}{c}_{\tau z_i z_j} ({\bf 0}|-\tau^{-1}) & = &\tau^2\,\overset{\star}{c}_{\tau z_i z_j} ({\bf 0}|\tau)\,,\\
\overset{\star}{c}_{\tau\tau z_i} ({\bf 0}|-\tau^{-1}) & = &\tau^3\,\overset{\star}{c}_{\tau\tau z_i} ({\bf 0}|\tau)\,,\\
\overset{\star}{c}_{\tau\tau\tau} ({\bf 0}|-\tau^{-1}) & = &\tau^4\,\overset{\star}{c}_{\tau\tau\tau} ({\bf 0}|\tau)\,.
\end{eqnarray*}
As $q\rightarrow 0$ (or equivalently, as $\tau\rightarrow i \infty$) the three structure functions vanish.
Hence they must be analytic in $q$ with vanishing constant term. This, with the modularity property,
implies they are cusp-forms of degrees 2, 3 and 4. However there are no no-zero cusp-forms of these
degree so
\begin{eqnarray*}
\overset{\star}{c}_{\tau z_i z_j} ({\bf 0}|\tau) & = & 0\,,\\
\overset{\star}{c}_{\tau\tau z_i} ({\bf 0}|\tau) & = & 0\,,\\
\overset{\star}{c}_{\tau\tau\tau} ({\bf 0}|\tau) & = & 0\,.
\end{eqnarray*}
Hence $A_{ij}=0$ and $B_i=0$ (ignoring quadratic terms in the prepotential) and $C$ is the triple
integral of the Eisenstein series $E_4$ which may be evaluted using the elliptic polylogarithm
function (see Appendix). Hence the result.
\end{proof}

\section*{Comments}

As in the Coxeter case, this basic example may be generalised in a number of different dirrections:

\begin{itemize}

\item{} By studying other Weyl groups. This should lead to dual prepotential akin to that in Theorem [\ref{maintheorem}]
(the $B_n$-case may be done very easily using the $A_{2n+1}$ to $B_n$ reduction).
A problem that this immediately generates is this: it follows from the general theory of
dual Frobenius manifolds that a prepotential exists and satisfies the WDVV equations. However
a direct verification would be considerably harder, presumably involving various
$\vartheta$-function identities and their modular counterparts. This should also
generate as a by-product elliptic Dunkl-type operators, since the flatness of the
dual Dubrovin connection is directly related, at least in the Coxeter case, to the
properties of the classical Dunkl operators.

\item{} The orbit space $\mathbb{C}\oplus{\mathfrak h}\oplus\mathbb{H}/J(A_n)$ is isomorphic to the
genus-one Hurwitz space $H_{1,n+1}(n+1)\,.$ One may easily
generalize to an arbitrary genus-one Hurwitz space
$H_{1,N}(k_1\,,\ldots\,,k_l)\,.$

\item{} More generally still, one may study the induced \lq dual\rq~structure on the
discriminant strata in these spaces.

\end{itemize}

Some of these points will be addressed in \cite{thesis} and others are under investigation. One
final questions is whether these elliptic solutions to the WDVV equations have any use in
Seiberg-Witten theory and Calogero-Moser
systems and their generalizations\cite{BMMM}.

\section*{Acknowledgments} Andrew Riley would like to thank the EPSRC for a research studentship. Ian Strachan
would like to thank the Erwin Schr\"odinger Institute, where part of the paper was written, for their
hospitality. Finally  we both would
like to thank Harry Braden, Misha Feigin and Sasha Veselov for various comments.

\appendix
\section{The integration of $\log\vartheta_1$ and elliptic polylogarithms}

The definition of the polylogarithm functions is, for positive
integers $N\,,$
\[
Li_N(z) = \sum_{r=1}^\infty \frac{z^r}{r^N}\,.
\]
The series converges for $|z|<1$, but the functions may be
analytically continued elsewhere as well as extended to
non-integer values of $N$ via integral representations of the
function. Immediate consequences of the definitions are:
\begin{eqnarray*}
Li_0(z) & = &\frac{z}{1-z}\,,\\
\frac{d~}{dz}Li_n(z) & = & \frac{1}{z} Li_{n-1}(z)\,.
\end{eqnarray*}
This function may be used to integrate $\log\vartheta_1(z)\,.$
Using the infinite product representation
\[
\vartheta_1(z|\tau) = 2 G q^{\frac{1}{4}} \sin z
\prod_{n=1}^\infty (1-q^{2n} e^{2iz})
\prod_{n=1}^\infty (1-q^{2n} e^{-2iz})
\]
(where $q=e^{i\pi\tau}$) one may write
\[
\log\vartheta_1(z|\tau)= \log(iGq^{\frac{1}{4}}) +
\left\{-iz-
\sum_{n=0}^\infty Li_1( q^{2n} e^{2iz})+
\sum_{n=1}^\infty Li_1( q^{2n} e^{-2iz})\right\}\,.
\]
In what follows it is convenient to define a new function
$\Lambda_N(z,q)$ as
\[
\Lambda_N(z,q) = -\frac{1}{2} \frac{(2iz)^N}{N!}-
\sum_{n=0}^\infty Li_N( q^{2n} e^{2iz})+(-1)^{N+1}
\sum_{n=1}^\infty Li_1( q^{2n} e^{-2iz})
\]
so
\[
{\frac{d~}{dz}} \Lambda_N(z,q) = 2 i \Lambda_{N-1}(z,q)
\]
and
\begin{eqnarray*}
\Lambda_0(z,q) & = & \frac{1}{2i} \frac{\vartheta'_1}{\vartheta_1}(z)\,, \\
\Lambda_1(z,q) & = & \log\vartheta_1(z) - \log(iGq^{\frac{1}{4}})\,.
\end{eqnarray*}
Recall the inversion formula
\[
(-1)^{N-1} Li_N(\zeta^{-1}) = Li_N(\zeta) + \sum_{j=0}^N \frac{B_j (2\pi \sqrt{-1})^j}{(N-j)!j!} (\log\zeta)^{N-j}
\]
where $B_j$ are the Bernoulli numbers.
From this it follows that
\begin{eqnarray*}
\Lambda_N(-z,q) +(-1)^N \Lambda_N(z,q) & = & (-1)^N
\sum_{j=1}^N \frac{B_j (2\pi \sqrt{-1})^j}{(N-j)!j!} (2iz)^{N-j}\,,\\
& = & {\rm polynomial~of~degree~}(N-1)\,.
\end{eqnarray*}
Remarkably, the function $\Lambda_N(z,q)$ is closely related to the
elliptic polylogarithm introduced by Beilinson and Levin \cite{BL,Levin}, following
the real-valued version introduced by Zagier\cite{Z}. By definition,
\[
\mathcal{L}i_r(q,\zeta) = \sum_{n=0}^\infty Li_r(q^n \zeta) +
\sum_{n=1}^\infty Li_r(q^n \zeta^{-1}) -
\chi_r(q,\zeta)\,,\qquad\quad r~{\rm odd}\,,
\]
where
\[
\chi_r(q,\zeta)=\sum_{n=0}^r \frac{B_{j+1}}{(r-j)!(j+1)!}
(\log\zeta)^{(r-j)} (\log q)^j\,.
\]
Hence, for example,
\[
\Lambda_3(z,q) = - \mathcal{L}i_3(q^2,e^{2iz}) + \frac{1}{3} (\log q) z^2 + \frac{1}{90} (\log q)^3\,.
\]
These formulae enable the equation
\[
\overset{\star}{c}_{ijk} = \sum {\rm
terms~involving~}\frac{\vartheta'_1}{\vartheta_1}
\]
to be integrate in closed form, leaving the answer in terms of the elliptic trilogarithmic
function. Note that these functions are multivalued, so have non-trivial monodromy group \cite{Ram}.
However such transformations change the elliptic trilogarithm by quadratic terms in the
flat $\{p^i\}$-variables and hence leave the physical structure functions unchanged.

It also follows from these formulae that
\begin{eqnarray*}
\left.\frac{\partial^3~}{\partial\tau^3} \mathcal{L}i_3(q^2,e^{2iz}) \right|_{z=0} & =& -\frac{i\pi^3}{15} E_4(\tau)\,,\\
\left.\frac{\partial^3~}{\partial\tau^2 \partial z} \mathcal{L}i_3(q^2,e^{2iz}) \right|_{z=0}  & =& 0\,,\\
\left.\frac{\partial^3~}{\partial\tau \partial z^2} \mathcal{L}i_3(q^2,e^{2iz}) \right|_{z=0} & =& \frac{2i \pi}{3} E_2(\tau)\,,
\end{eqnarray*}
where $E_2(\tau)\,,E_4(\tau)$ are the normalised Eisenstein series.


\begin{thebibliography}{99}

\bibitem{BL} Beilinson, A. and Levin, A.,{\sl The Elliptic Polylogarithm\,,} in {\sl Motives}
(ed. Jannsen, U., Kleiman, S,. Serre, J.-P.), Proc. Symp. Pure Math. vol {\bf 55}, Amer. Math. Soc.,
(1994), Part 2, 123-190.



\bibitem{B} {Bertola, M. {\sl Frobenius manifold structure on orbit space of Jacobi groups; Parts I and II},
Diff. Geom. Appl. {\bf 13}, (2000), 19-41 and  {\bf 13}, (2000), 213-23.}

\bibitem{BMMM} {Braden, H.W., Marshakov, A., Mironov, A. and Morozov, A.,
{\sl Seiberg-Witten Theory for a Nontrivial Compactification from Five-Dimensions to Four-Dimensions,}
Phys. Lett. {\bf B448}, 195-202, 1999.}

\bibitem{ChalVes} {Chalykh, O. and Veselov, A.P., {\sl Locus configurations and $\vee$-systems,}
Phys. Lett. {\bf A285} 339-349, 2001.}


\bibitem{dubrovin1} {Dubrovin, B., {\sl Geometry of 2D topological field theories} in {\sl Integrable
Systems and Quantum Groups}, ed. Francaviglia, M. and Greco, S.. Springer lecture
notes in mathematics, {\bf 1620}, 120-348.}

\bibitem{dubrovin2} Dubrovin, B., {\em On almost duality for Frobenius manifolds} in
{\sl Geometry, topology, and mathematical physics}, 75--132,
Amer. Math. Soc. Transl. Ser. 2, 212,
Amer. Math. Soc., Providence, RI, 2004.

\bibitem{EZ} Eichler, M. and Zagier, D., {\sl The Theory of Jacobi Forms}, {Birkh\"auser, 1985}
({\sl Progress in Mathematics} Vol. 55).

\bibitem{FV} Feigin, M. and Veselov, A.P., {\sl Private communication} and talk given at
SISSA, Sept. 2005.

\bibitem{Kawai} Kawai, T., {\sl String duality and enumeration of curves by
Jacobi forms} in {\sl
Integrable systems and algebraic geometry} (Kobe/Kyoto, 1997), 282--314,
World Sci. Publishing, River Edge, NJ, 1998.


\bibitem{Levin} Levin, A., {\sl Elliptic polylogarithms: an analytic theory}
Compositio Math. 106 (1997), no. 3, 267--282.

\bibitem{Marsh} Marshakov, A., Mironov, A., and Morozov, A., {\sl A WDVV-like equation in
$N=2$ SUSY Yang-Mills theory}, Phys. Lett. {\bf B 389} (1996) 43-52.

\bibitem{iabs} Strachan, I.A.B., {\em Frobenius manifolds: natural
submanifolds and induced bi-Hamiltonian structures}, Differential
Geometry and its Applications, 20 (2004), 67-99.

\bibitem{Ram} Ramakrishnan, D., {\sl On the monodromy of higher logarithms},
Proc. Amer. Math. Soc. 85 (1982), no. 4, 596--599.

\bibitem{thesis} Riley, A., {\sl} Ph.D. thesis, Hull University, 2006.

\bibitem{wirth} Wirthm\"uller, K., {\sl Root systems and Jacobi forms}, Compositio Mathematica {\bf 82}
(1992) 293-354.

\bibitem{W} Whittaker, E.T. and Watson, G.N., {\sl A course of Modern Analysis}, Cambridge University
Press (any edition).

\bibitem{Z} Zagier, D. {\sl The Bloch-Wigner-Ramakrishnan polylogarithm function},
Math. Ann. {\bf 286} (1990) 613-624.

\end{thebibliography}
\end{document}